\documentstyle[11pt,aaspp4]{article}     

\slugcomment{To be published in the {\it Astronomical Journal}, July 1999 issue}
\lefthead{Courteau and van den Bergh}
\righthead{The Solar Motion Relative to the Local Group}

\begin{document}



\font\hugemath=cmsy10 scaled \magstep1
\def\Sum{{\hugemath\Sigma}}


\def \Rx        {R_{exp}}
\def \V22{V_{2.2}}
\def \r22{R_{disk}}
\def \partialvr {\partial\log{\V22} \thinspace / \thinspace\partial\log{\Rx}}
\def \Br        {B-r}
\def \BI        {B-I}
\def \Mr        {M_r}
\def \MI        {M_I}
\def \Reff      {R_{eff}}
\def \rc        {r_{\rm c}}
\def \gtorder{\mathrel{\raise.3ex\hbox{$>$}\mkern-14mu
             \lower0.6ex\hbox{$\sim$}}}
\def \ltorder{\mathrel{\raise.3ex\hbox{$<$}\mkern-14mu
             \lower0.6ex\hbox{$\sim$}}}
\def\degd#1.#2{                         
               \ifmmode {#1^{\hskip 0.05em\circ}\hskip-0.42em.\hskip0.08em#2}
               \else {#1$^{\hskip 0.05em\circ}\hskip-0.42em.\hskip0.08em$#2}
               \fi
              }
\def \reminder#1{\vskip 0.5truecm\noindent{\sl #1}}
\def \xx{{\bf ????}}
\def \etal {{\it et al.~}}
\def \eg {{\it e.g.,~}}
\def \ie {{\it i.e.,~}}
\def \vs {{\it vs~}}
\def \cf {{cf.}}
\def \o{\overline}
\def \u{\underline}
\def \t{\tilde}

\def\asec{^{\prime\prime}}
\def\kms{\ifmmode {\rm \, km \, s^{-1}} \else $\rm \,km \, s^{-1}$\fi}  
\def\ksmpc{\ifmmode{\rm km}\,{\rm s}^{-1}\,{\rm Mpc}^{-1}\else km$\,$s$^{-1}\,$Mpc$^{-1}$\fi}
\def\pc{{\rm\,pc}}
\def\kpc{{\rm\,kpc}}
\def\Mpc{{\rm\,Mpc}}
\def\dnsigma  {$D_n$-$\sigma$}
\def\sol{$_\odot$}
\def\sun{$_{\scriptscriptstyle\odot}$}
\def\solar{\ifmmode_{\mathord\odot}\else$_{\mathord\odot}$\fi}  
\def\lsol     {$L_\odot$}
\def\msol{\ifmmode M_\odot\else$M_\odot$\fi}
\def\Msol{\ifmmode M_\odot\else$M_\odot$\fi}
\def\hunit{km~sec$^{\hbox{\scriptsize -1}}$~Mpc$^{\hbox{\scriptsize -1}}$~}

\def\onesigma{$1$-$\sigma$}
\def\dline{\noalign{\hrule height1pt\vskip 2pt\hrule height 1pt\vskip 4pt}}
\def\bline{\noalign{\vskip 5pt\hrule height1pt}}
\def\pp{\par\yskip\noindent\hangindent 0.4in \hangafter 1}
\def\reference#1#2#3#4 {\pp#1, {\it#2}, {\bf#3}, #4.}
\def\yskip{\penalty-50\vskip3pt plus3pt minus2pt}
\def\yyskip{\penalty-100\vskip6pt plus6pt minus4pt}
\def \littlemm{\ifmmode{\scriptscriptstyle m }
     \else{\hbox{$\scriptscriptstyle m $ }}\fi}
\def \topemm{\raise .9ex \hbox{\littlemm}}
\def \magpoint{\hbox to 2pt{}\rlap{\hskip -.5ex \topemm}.\hbox to 2pt{}}
\def\deg {$^\circ$}
\def\fr#1#2{{#1\over#2}}
\def\onetwo{{\textstyle {1 \over 2} \displaystyle}}
\def\threetwo{{\textstyle {3 \over 2} \displaystyle}}
\def\alphatwo{{\textstyle {\alpha \over 2} \displaystyle}}
\def\quarter{{1\over 4}}
\def\piby2{{\pi \over 2}}
\def\threeh{{3\over 2}}
\def\fiveh{{5\over 2}}
\def\d{{\rm d}}
\def\VP{{\rm VP}}


\def\spose#1{\hbox to 0pt{#1\hss}}
\def\lta{\mathrel{\spose{\lower 3pt\hbox{$\sim$}}
    \raise 2.0pt\hbox{$<$}}}
\def\gta{\mathrel{\spose{\lower 3pt\hbox{$\sim$}}
    \raise 2.0pt\hbox{$>$}}}

\def\hub      {$H_{\hbox{\scriptsize 0}}$}
\def\hunit    {\kms~Mpc$^{-1}$}
\def\chisqr   {$\chi^2$}
\def\chidof   {$\chi^2_\nu$}
\def\ion#1#2{#1$\;${\small\rm{#2}}\relax}
\def\halpha   {{H$\alpha$\ }}
\def\ha       {H$\alpha$\ }
\def\hi       {\ion{H}{I}\ }
\def\hii      {\ion{H}{II}\ }
\def\atcenter#1{\hfil#1\hfil}


\title{The Solar Motion Relative to the Local Group}

\author{St\'ephane Courteau and Sidney van den Bergh}

\affil{National Research Council, Herzberg Institute 
       of Astrophysics}
\affil{Dominion Astrophysical Observatory}
\affil{5071 W. Saanich Rd, Victoria, BC \ V8X 4M6 \ Canada}


\begin{abstract}

New data on the membership of the Local Group, in conjunction with new
and improved radial velocity data, are used to refine the derivation 
of the motion of the Sun relative to the Local Group (hereafter LG).  
The Sun is found to be moving with a velocity of $V = 306\pm 18$ \kms\ 
towards an apex at $\ell= 99^\circ \pm 5^\circ$ and $b = -4^\circ \pm 4^\circ$. 
This agrees very well with previous analyses, but we discuss the possibility
of a bias if the phase-space distribution of LG galaxies is bimodal. 
The LG radial velocity dispersion is $61 \pm 8$ \kms.
We use various mass estimators to compute the mass of the Local Group and 
the Andromeda subgroup.  We find $M_{LG}=(2.3\pm0.6) \times 10^{12}$ 
\msol, from which $M/L_V = 44 \pm 12$ (in solar units).  For an assumed LG age 
of $14 \pm 2$ Gyr, the radius of an idealized LG zero-velocity surface is
$r_\circ= 1.18 \pm 0.15$ Mpc.  The Local Group is found to have 35 likely 
members.  Only three of those have (uncertain) distances $\gta 1.0$ Mpc 
from the LG barycenter.  Barring new discoveries of low surface brightness
dwarfs, this suggests that the Local Group is more compact, and isolated 
from its surroundings, than previously believed.

\end{abstract}





\section{Introduction}
 
The motion of the Sun relative to the other members of the Local Group
has been been studied for many decades, with investigations by 
Humason, Mayall \& Sandage (1956), Yahil, Tammann \& Sandage (1977; hereafter
YTS77), Sandage (1987), Karachentsev \& Makarov (1996), and 
Rauzy \& Gurzadyan (1998 and references therein; hereafter RG98).  YTS77 and 
many others have stressed the
importance of understanding the solar motion relative to the Local
Group in the context of large-scale motions of galaxies.  The reliability of
measurements of peculiar motions in the Universe, or residual motion
from the uniform Hubble expansion, depends, in part, on accurate knowledge
of the motion of the solar system relative to any standard inertial
frame.  This inertial rest frame is usually taken as the centroid of 
the Local Group of galaxies, or the reference frame in which the dipole
of the Cosmic Microwave Background (CMB) vanishes (Kogut \etal 1993).  
The motion of the Sun 
relative to the cosmic microwave background can be decomposed into
into a sum of local and external components\footnote{Our notation is 
analogous to that of RG98}:

\begin{equation}
V_{{\rm Sun} \rightarrow {\rm CMB}} = V_{{\rm Sun} \rightarrow {\rm LSR}} + 
                                      V_{{\rm LSR} \rightarrow {\rm GSR}} + 
                                      V_{{\rm GSR} \rightarrow {\rm LG }} +
                                      V_{{\rm LG}  \rightarrow {\rm CMB}}.
\end{equation}

$V_{{\rm Sun} \rightarrow {\rm LSR}}$ is the motion of the Sun relative 
to the nearby stars which define a Local Standard of Rest, and the motion 
$V_{{\rm LSR} \rightarrow {\rm GSR}}$ is the circular rotation of the LSR 
about the Galactic center that is directed towards $\ell =90^\circ$ and 
$b=0^\circ$.  Externally, $V_{{\rm GSR}  \rightarrow {\rm LG}}$ is the 
motion of the Galactic center (or Galactic Standard of Rest) relative to 
the LG centroid, which is caused by non-linear dynamics within 
the Local Group (mostly infall of the Galaxy toward M31).
Finally, $V_{{\rm LG} \rightarrow {\rm CMB}}$ is the peculiar velocity of the 
Local Group in the CMB rest frame, induced by gravitational perturbations in 
the Universe.  

Recent discoveries of new candidate members of the Local Group, and deletion
of former candidates, have allowed us to revise the solution for the solar 
motion relative to the Local Group, $V_{{\rm Sun} \rightarrow {\rm LG}}$,
assess Group membership, and compute a new value for the mass of the Local Group.  
Three criteria are usually invoked to assess the probability that a 
galaxy is associated with the Local Group: (1) The distance to that galaxy 
from the LG barycenter should be less than (or comparable to) the radius 
at the zero-velocity surface (Lynden-Bell 1981, Sandage 1986),
(2) it should lie close to the ridge-line solution between radial 
velocity and the cosine of the angle from the solar apex relative to
well-established Local Group members, and (3), it should not appear to 
be associated with any more distant group of galaxies that is centered 
well beyond the limits of the Local Group.  We examine these criteria 
below.  

This paper is organized as follows.  First, we compute a new solution
for the motion of the Sun relative to LG members in \S 3.  We then estimate 
the radius of the zero-velocity surface in \S 4, and assess Local Group 
membership in \S 5, based on the three criteria listed above.  We conclude in \S 6 
with a brief discussion and summary, and a digression on the detectability
of small groups like the LG using X-ray telescopes.

Recent studies of the membership in the Local Group also include 
van den Bergh (1994a,b), Grebel (1997), and Mateo (1998).  The reader 
is referred to van den Bergh (2000; hereafter vdB2000) for a comprehensive 
review on the nature of, and membership in, the Local Group.  

\section{The Data}

  
A listing of information on the 32 probable (+3 possible) members of the 
Local Group, that were isolated using the criteria discussed above, is given 
in Table 1.  Columns (1-3) give the names and David Dunlap Observatory 
morphological types (van den Bergh 1966, 1994b) for each Local Group member. 
Equatorial (J2000) and Galactic coordinates are listed in columns (4-7).  
Various photometric parameters (visual color excess, absolute visual 
magnitude, and distance modulus) taken from vdB2000 are listed in columns (8-10).  
Column (11) gives the heliocentric radial velocity of each galaxy in \kms, 
and column (12) 
lists the cosine of the angle between each galaxy and the solar motion apex 
in the rest frame of the Local Group.
Columns (13-14) give the distance of a galaxy from the Sun and from the 
LG barycenter in Mpc.  Finally, column (15) gives the main reference to
each of the radial velocities quoted in column (11). 
The Local Group suspects at large distances, marked with an asterisk
in Table 1, are Aquarius (=DDO 210) with a distance from the LG center
of $1.02 \pm 0.05$ Mpc (Lee 1999), Tucana at $\simeq 1.10 \pm 0.06$ Mpc 
(vdB2000), 
and SagDIG with a poorly determined LG distance of $1.20 \lta D \lta 1.58$ 
(Cook 1987).  Uncertain entries in Table 1 are followed by a colon. 

The positions of Local Group members in cartesian Galactocentric 
coordinates are shown in Fig.~1.  The velocity components, X, Y, Z,
of an object point toward the Galactic center ($\ell=0^\circ,
b=0^\circ$), the direction of Galactic rotation ($\ell=90^\circ,
b=0^\circ$), and the North Galactic Pole ($b=90^\circ$).  


We have calculated distances of individual galaxies, relative to the 
Local Group barycenter, by (1) assuming that most of the LG mass is 
concentrated in the Andromeda and Galactic subgroups, (2) adopting 
a distance to M31 of 760 kpc (vdB2000), and (3) assuming that M31 
is 1.5 more massive than the Milky Way (Mateo 1998, Zaritsky 1999, 
and references therein).  
Lacking more detailed information about the mass constituents of
the Local Group, it seems reasonable to expect that the local center 
of mass will be situated on the line between our Galaxy and M31 in the 
direction of M31.  The Local Group barycenter is located at 
0.6 times the distance to M31 at 454 kpc toward $\ell=\degd 121.7$
and $b=\degd -21.3$.  This corresponds, in Galactic cartesian 
coordinates, to X=$-220$, Y$=+361$, and Z$=-166$ kpc. 



Histograms of the differential and cumulative distance distributions of the 
LG members, relative to its barycenter, are shown in Figs. 2 and 3, 
respectively.  These figures show that all probable LG members have distances
$\lta 850$ kpc from the dynamical center of the LG.  Taken at face value, 
this suggests that the core of the Local Group may be smaller, and more 
isolated from the field, than has generally been assumed previously 
(\eg Jergen, Freeman, \& Bingelli 1998, Pritchet 1998).

\section{Solar Motion Relative to Local Group Members}

Using the line-of-sight velocities and positions of probable Local Group 
members (Table 1), we compute a new solution for the bulk motion of the Sun 
relative to the Local Group centroid.  The computation of a bulk flow, 
${\boldmath v^B}$, is independent of estimated distances to any of the
galaxies, or the exact shape of their orbits, provided the spatial and 
velocity distributions are independent (\eg YTS77, RG98).  
If the 3-dimensional velocity distribution is invariant under spatial 
translations, one can further assume that the global velocity field 
can be decomposed into the sum of a bulk flow ${\boldmath v^B}$ and a 
3-dimensional random isotropic Maxwellian distribution with a velocity 
dispersion $\sigma_v$.  The bulk flow statistics reduces to the 
maximisation of the likelihood function, 

\begin{equation}
{\cal L} = - \ln \sigma_v - \frac{1}{N} \sum_{k=1}^{N}
 \frac{( v^k_r - v^B_x \hat{r}^k_1 - v^B_y \hat{r}^k_2 
       - v^B_z \hat{r}^k_3)^2}{2\sigma_v^2}, 
\end{equation}

where $v^k_r$ is the observed radial velocity of galaxy $k$, and the 
components $\{ {\hat{r}^k_j} \}_{j=1,3}$ are the direction cosines of 
that galaxy.  The inferred solar apex corresponds to the direction which 
minimizes the scatter in the distribution of radial velocities versus 
$cos \theta$, where $\theta$ is the angle between the solar apex and the unit 
vector towards each galaxy.  

Details on such techniques, and confidence in the estimators, can be 
found in YTS77 and RG98 (our estimator is identical to that developed 
by RG98.)  The observational errors in the radial velocities are 
relatively small, and insignificant compared with the residual velocity 
dispersion.  They are therefore neglected. 
Here we adopt the values quoted by the main source in Col. (15) of Table 1. 
Standard deviations for the amplitude and direction of the solar
motion, and for the residual velocity dispersion of the Local Group, are
estimated by bootstrap resampling of the input data.  The errors quoted 
correspond to the $1\sigma$ dispersion for each parameter.  

A maximum likelihood solution, giving equal weight to all 26 objects with
measured heliocentric radial velocities, yields a solar motion with 
V\sol = $306\pm18$ \kms\ towards an apex 
at $\ell= 99^\circ \pm 5^\circ$ and $b = -3^\circ \pm 4^\circ$.  
The residual radial velocity dispersion in the Local Group is 
$\sigma_r = 61 \pm 8$ \kms.  
Assuming the velocity distribution of Local Group galaxies to 
be isotropic, the three-dimensional velocity dispersion in 
the Local Group is $\simeq 106$ \kms.  

\begin{table}[h]
\centering
\tablenum{2}
\begin{tabular}{rrrl}
\multicolumn{4}{c}{Table 2} \\
\multicolumn{4}{c}{Solar motion relative to the Local Group} \\ [2pt] \hline\hline
 V\sol & \atcenter{$\ell$} & \atcenter{$b$} & Reference \\
(\kms) & \atcenter{$(^\circ)$} & \atcenter{$(^\circ)$} & \\ [4pt] \hline 
   & & & \\ [-10pt]
 $306\pm \phantom{0}18$ & $ 99\pm\phantom{4}5$ & $ -3\pm\phantom{2}4$ & This paper  \\
 $305\pm           136$ & $ 94\pm          48$ & $-34\pm          29$ & RG98        \\
 $316\pm \phantom{00}5$ & $ 93\pm\phantom{4}2$ & $ -4\pm\phantom{2}2$ & Karachentsev \&   \\
                        &                      &                      & Makarov (1996)    \\
  295\hspace{29pt}      &   97\hspace{24pt}    & $ -6$\hspace{24.5pt} & Sandage (1986)    \\
 $308\pm \phantom{0}23$ & $105\pm\phantom{4}5$ & $ -7\pm\phantom{2}4$ & YTS77             \\
 $315\pm \phantom{0}15$ & $ 95\pm\phantom{4}6$ & $ -8\pm\phantom{2}3$ & de Vaucouleurs \& \\
                        &                      &                      & Peters (1968)     \\ 
 $292\pm \phantom{0}32$ & $106\pm\phantom{4}6$ & $ -7\pm\phantom{2}4$ & Humason \&        \\
                        &                      &                      & Wahlquist (1955)  \\ 
 $308\pm \phantom{0}26$ & $ 93\pm\phantom{4}6$ & $-14\pm\phantom{2}4$ & Mayall (1946) \\ [4pt] \hline

\end{tabular}
\end{table}


A comparison with other published solutions for the
motion of the Sun relative to the Local Group barycenter is given 
in Table 2. 
With the exception of RG98, most solutions are found to be in agreement 
with each other to within their quoted errors.  For example, our solution 
seldom differs by more than 1 \kms\ from YTS77, with a maximum deviation 
of $\pm2$ \kms.  
The good agreement among most published solutions is not fortuitous.  
Local Group dynamics are heavily dominated 
by systems that were already included in the sample of Mayall (1946; the earliest
reference cited here).  Addition of new members, especially to the Galaxy subgroup
(which now accounts for half of all known LG members with a measured redshift), 
has not altered the solar motion solution in any significant way.  Moreover, 
most studies of solar motion relative to LG
galaxies have assumed a uniform potential that governs LG dynamics. 
The solar motion amplitude measured by Sandage (1986) assumes a two-to-one
mass ratio between M31 and our Galaxy.  A lower mass ratio of 1.5, as we 
advocate here, would yield an even lower amplitude.
The result by RG98 differs more substantially from all others due to the 
different nature of their approach.  As a first step, RG98 recognize the 
existence of the two main dynamical substructures within the LG, namely the 
Galaxy subgroup (13 galaxies\footnote{RG98's ``Milky Way'' subgroup does 
not include the newly discovered Sagittarius dwarf spheroidal.}) and the 
Andromeda subgroup (7 galaxies).  For each subgroup, RG98 estimate a bulk 
flow using Eq. (2).  They find, in Galactocentric coordinates:
$V_{Galsub \rightarrow Sun}=(94 \pm 64,-354\pm 42,37 \pm33)$ \kms\
or $|V_{Galsub \rightarrow Sun}|=368 \pm 28$ km s$^{-1}$ toward 
$(\ell=285^\circ \pm 11^\circ,b=+6^\circ \pm 5^\circ)$, 
and $V_{Andsub \rightarrow Sun}=( -127 \pm 541,-143\pm 267,301 \pm254)$ 
\kms\ or $|V_{Andsub \rightarrow Sun}|=357 \pm 218$ km s$^{-1}$ 
toward ($\ell=228^\circ \pm 180^\circ, b=+58^\circ \pm 65^\circ$). 
The error bars for the bulk flow estimate of the Andromeda subgroup
are large, due to 
the small number of galaxies involved in the statistics, and because of the 
narrow angular size of the subgroup on the sky (\ie bulk flow components 
perpendicular to the M31 line-of-sight are poorly constrained.)
RG98 compute the global bulk flow of the LG as the mean motion of its 
main dynamical substructures, equally weighted, \ie
$V_{LG \rightarrow Sun}=(V_{Galsub \rightarrow Sun}
+V_{Andsub \rightarrow Sun})/2$, which gives
$V_{LG \rightarrow Sun}=(-17 \pm 303, -249\pm 155, 169 \pm144)$
or $|V_{LG \rightarrow Sun}|=301$ km $s^{-1}$ toward $(l=266,b=+34)$. 
The residual velocity dispersion is $\sigma_r = 110.3$ \kms.
Error bars are thus larger in RG98's treatment because of the poor estimate 
of the M31 subgroup's bulk flow. 

RG98 suggest that the phase-space distribution of LG galaxies is bimodal.
Application of bulk flow statistics from Eq. (2) to a uniform LG 
distribution may therefore be biased.  Indeed, with the exception of RG98, 
the results quoted in Table 2 apply if the velocity distribution function 
of selected LG galaxies is the sum of a 3-dimensional bulk flow, plus a 
random component that does not correlate with the spatial position of 
galaxies.  However, solar motion solutions that assume a uniform 3D 
structure for the Local Group may be biased if the Andromeda subgroup 
bulk flow $V_{Andsub \rightarrow Sun}$ differs significantly from that 
of the Galaxy subgroup $V_{Galsub \rightarrow Sun}$.  
This suggestion is supported by RG98's analysis and corroborated by our 
own reexamination of this issue.  Our analysis, based on Eq. (2), also 
suggests that the Andromeda subgroup would partake of a different, stronger, 
bulk motion than the Galaxy subgroup.  But it would be premature to make 
any claims based on these results, given the large errors in the apex 
parameters of the Andromeda subgroup.  In any case, the poor number 
statistics do not allow a rejection, or confirmation, of this hypothesis.  
All solutions, which account for subgrouping or a uniform structure of
the Local Group, agree to within their $1\sigma$ confidence interval.  

\subsection{A summary of corrections to radial velocities}

The correction to heliocentric radial velocities, $V_{\rm hel}$, for a solar 
apex of direction ($\ell_a$, $b_a$), and amplitude $V_a$ in any reference 
frame can be expressed as: 

\begin{equation}
V_{corr} = V_{\rm hel} + V_a \left(\cos b \cos b_a \cos (\ell-\ell_a) +
           \sin b \sin b_a\right), 
\end{equation}


\noindent{where $\ell$, and $b$ are the Galactic coordinates to the observed 
galaxy.}  The peculiar motion of the Sun relative to the Local Standard of Rest
(LSR) is 16.5 \kms\ towards $\ell = 53^\circ$ and $b = +25^\circ$ (Delhaye 
1965; see also Crampton 1968), or X$=+9$, Y$=+12$, and Z$=+7$ 
\footnote{Note the typographical error in Eq. (1) of Braun \& Burton 
(1998).} \kms.  Therefore, 
\begin{equation}
V_{\rm LSR} = V_{\rm hel} + 9\cos \ell \cos b + 12\sin \ell \cos b + 7\sin b
\end{equation} 

The Galactic rotation has an amplitude Y$=220\pm20$ \kms (X=0, Z=0) 
toward $\ell = 90^\circ$ and $b = +0^\circ$ (IAU 1985 convention;
see Kerr and Lynden-Bell 1986).
Therefore, the corrected radial velocity of a galaxy in the Galactic 
Standard of Rest is: 

\begin{equation}
V_{\rm GSR} = V_{\rm hel} + 9\cos \ell \cos b + 232 \sin \ell \cos b + 7\sin b.
\end{equation} 

The corrections given above are widely used and accepted (RC3; de Vaucouleurs
\etal 1991).  We also found, in \S 3.0, the correction for motion of the Sun relative 
to the Local Group centroid as:

\begin{equation}
V_{\rm LG}({\rm this~paper}) = V_{\rm hel} - 79 \cos \ell \cos b 
                               + 296 \sin \ell \cos b - 36 \sin b,
\end{equation}

under the assumption that the velocity distribution function in the Local Group
can be described as bulk flow plus a random isotropic Maxwellian component. 
Applying the same premises, but to the two main LG substructures instead of
the LG as a whole as we did, RG98 find (also Table 2),

\begin{equation}
V_{\rm LG} ({\rm RG98}) = V_{\rm hel} - 18 \cos \ell \cos b 
                          + 252 \sin \ell \cos b - 171 \sin b.
\end{equation}

The RC3 does not include any corrections for galaxy motions in the frame 
of the 
Local Group, on account of their ill-defined nature.  This was perhaps a wise 
decision.  The RC2 (de Vaucouleurs \etal 1976) reports the ``old'' solar apex 
solution ($300\sin l\cos b$), but modern solutions (Table 2) show deviations 
from the RC2 formulation as large as $\pm87$ \kms, as already noted by YTS77.
Perhaps even more important are the deviations that exist between our
solution and RG98.  The maximum deviations (Eqs. 6-8) are $\pm154$ \kms\ 
toward ($\ell=145^\circ$, $b=60^\circ$) and ($\ell=325^\circ$, 
$b=-60^\circ$).  These are shown in Figure 4; negative and positive 
residuals are represented by stars and circles, respectively. 

In choosing a reference frame for cosmological studies, one may transform 
heliocentric radial velocities to the CMB frame (e.g. Kogut \etal 1993).
Under the assumption that the CMB dipole is kinematic in origin, and not
due to any external force field, this operation carries little uncertainty.
On the other hand, the transformation to the Local Group rest frame is free 
of any assumptions about the origin of the CMB dipole, and minimizes the
effect of the mass distributed outside the sample. 
Modern solutions for solar motion with respect to LG galaxies that assume 
a uniform LG potential, and a fixed LG barycenter, are robust.  These studies 
yield nearly identical solutions (\eg Table 2).  However, this result is 
either due to the regular nature of the LG or to the fact that we are making 
similar erroneous assumptions.  The kinematical method described above can
lead to biased results if the phase-space galaxy distribution is not 
homogenous.  Use of a dynamical method to reconstruct the orbits of 
individual LG galaxies could provide a potentially more accurate 
description of the motion of the LG center of mass.  Such a method based 
on Least-Action principles has been proposed (Shaya, Peebles, \& Tully 
1995), but it depends on a reliable knowledge of the galaxy distribution 
outside the Local Group, which is lacking at present. 
Clearly, it is the prerogative of the astronomer to adopt (and justify)
whatever cosmological rest frame he/she prefers.  

Given that the mean motion of the Local Group is consistent
with the combined motion of its two main substructures, we will adopt the 
``standard'' solution [Eq.(6)] as the best description for solar motion 
relative to the Local Group.  However, one should keep in mind 
the main caveats/assumptions for this solution, as we reiterate in \S 6.

\section{Mass of the Local Group}

  If we assume that the LG is in virial equilibrium, and that its 
velocity ellipsoid is isotropic ($\sigma^2 = 3\sigma_r^2$), then the 
mass of the Local Group can be computed from its velocity dispersion 
as (Spitzer 1969; see also Binney \& Tremaine 1987, eq. 4-80b):

\begin{equation}
 M_{LG} \ \simeq \ \frac{7.5}{G} \ \langle \sigma_r^2 \rangle \ r_h \ = 
            \ 1.74 \times 10^6 \ \langle \sigma_r^2 \rangle \ r_h 
            \ \ \msol,
\end{equation}


where $r_h$ is the radius in kpc containing half the mass, as measured from 
the center of the isotropic distribution.  The numerical value of $r_h$ can 
be estimated from the cumulative distance distribution
of LG members shown in Figure 2.  We find that $r_h \ltorder 450$ kpc.  
This number-weighted figure is clearly an upper limit to the actual 
mass-weighted estimate.  If M31 accounts for $\sim 60\%$ of the mass 
in the Local Group, 
a simple mass distribution model gives $r_h \simeq 350$ kpc. 
Using this value and $\sigma_r = 61 \pm 8$ \kms, we find 
$M_{LG} = (2.3 \pm 0.6) \times 10^{12} \msol. $

It is of interest to tally the mass of individual LG components. 
To compute the mass of the Andromeda subgroup, we use the projected mass
method of Bahcall \& Tremaine (1981) and Heisler \etal\ (1985; see
also Aceves \& Perea 1999).  In the
absence of specific information on the distribution of orbital 
eccentricities, the projected mass estimator is given by:

\begin{equation}
 M_{PM} = \frac{10.2}{G(N-1.5)} \ \sum_i V^2_{zi} R_i, 
\end{equation}

where $R$ is the projected separation from M31 (assuming $D_{M31}=760$ kpc),
and $V^2_z$ is the radial velocity in the frame of M31. 
Table 3 gives the relevant parameters for all 7 known members of 
the Andromeda subgroup.
We find that the Andromeda subgroup has a mass of 
$(13.3\pm1.8) \times 10^{11}$ \msol, the lower and upper 
bounds corresponding to the virial and projected mass estimates, 
respectively, following the notation of Heisler \etal (1985).

From the inferred motion of nearby satellites, Zaritsky (1999) shows 
that the Galactic subgroup has a mass of $(8.6\pm4.0) \times 10^{11} \msol.$  
Thus, the two major subgroups have a combined mass of 
$(21.9\pm4.4) \times 10^{11} \msol.$
This may be compared to the virial mass of $(23\pm6)\times 10^{11}$\msol\
found above for the entire LG.  This agreement may be fortuitous if the 
LG is not in virial equilibrium or if the LG potential is non-isotropic. 
However, taken at face value, this result suggests that most of the 
dark and luminous mass in the Local Group is locked into the Andromeda
and Galactic subgroups, unless the intra-cluster dark matter is distributed 
in a highly flattened shape. 
The timing argument by Kahn and Woltjer (1959), which is based on the motion 
of M31 towards the Galaxy yields a minimum Local Group mass of 
$\sim 18 \times 10^{11}$ \msol.  Sandage (1986), using a similar argument
for the deceleration of nearby galaxies caused by the Local Group, finds 
a maximum mass for the Local Group equal 
to $5 \times 10^{12}$ \msol, with a best-fit value of $4 \times 10^{11}$ \msol.
He also arrives at this low value by using the dispersion as a virial velocity 
to compute a virial mass for the Local Group.  The formula he used for the 
virial mass differs by a factor 7.5 from ours (Eq. 8), introduced by 
replacing $\sigma^2 = 3\sigma_r^2$ for an isotropic velocity ellipsoid
and considering the half-mass radius, $r_h$, instead of the ill-defined 
gravitational radius $r_g$.   Sandage also used an estimate for $r_g$
that is too small by a factor $\sim 2$ (if $r_h \simeq 0.4 r_g$). 
This explains the discrepancy ``by a factor of 7'' (with Kahn-Woltjer)
discussed by Sandage. 
Moreover, his result that $M_{\rm LG} = 4 \times 10^{11}$ \msol\ based 
on a velocity perturbation analysis of the Local Group, assumes a
formation age of 18.1 Gyr (\hub=55 \hunit\ for an $\Omega=0$ Universe),
and that $M_{\rm M31}$ = 2 $M_{\rm Gal}$.  Adoption of revised figures, 
\hub=65 \hunit\ and $M_{\rm M31}$ = 1.5 M$_{\rm Gal}$, yield a model-data 
comparison that agrees perfectly well with 
$M_{LG} = (2.3 \pm 0.6) \times 10^{12}$ \msol\ 
(see Sandage 1986, fig. 11).
Thus, both calculations in Sandage (1986) are consistent with a
higher value for M$_{\rm LG}$, equal to the one we measure.

\begin{table}[t]
\centering
\tablenum{3}
\begin{tabular}{lrlrc}
\multicolumn{5}{c}{Table 3} \\
\multicolumn{5}{c}{The Andromeda Sub-Group} \\ [2pt] \hline\hline
 Name & $R^a$ & $v_{\rm hel}$ &
 $v_{\rm cor}^b$ & \atcenter{$q^c/10^{11} \msol$} \\ [4pt] \hline
   & & & & \\ [-10pt]
 M 32    &   5.3 & -205 &  95 & 0.11 \\
 NGC 205 &   8.0 & -244 &  58 & 0.06 \\
 NGC 185 &  93.9 & -202 & 107 & 2.49 \\
 NGC 147 &  98.3 & -193 & 118 & 3.18 \\
 M 33    & 197.3 & -181 &  72 & 2.37 \\
 IC 10   & 242.9 & -344 & -29 & 0.48 \\
 Pisces  & 263.0 & -286 & -38 & 0.90 \\ [4pt] \hline
\end{tabular}

\caption[test]{(a) The projected separations in kpc are based on a 
  distance to M31 of 760 kpc.  Compare with Bahcall \& Tremaine 1981, 
  Table 4.; 
\noindent (b) The velocities $v_{\rm cor}$ are corrected for the solar
 motion relative to the Local and Galactic Standards of Rest (Eq. 5), and 
 for radial motion of the Galaxy toward M31, i.e. $v_{\rm cor} = V_{GSR} +
 124 (\cos b \cos(-21.3) \cos (\ell-121.7) + \sin b \sin(-21.3)$;
\noindent (c) The projected mass $q \equiv v^2_z R/G \ \msol$. }
\end{table}

From the absolute magnitudes of LG galaxies listed in Table 1, we 
compute the total luminosity of the Local Group to be
$L_V = 5.2 \times 10^{10} L_\odot$\footnote{Adopting
$M_{V\odot} = +4.82 \pm 0.02$ (Hayes 1983).}, corresponding to 
$M_V (LG) = -22.0.$  Combined with our estimate of the virial mass, 
and assuming a 10\% error in $L_V$, we measure $M/L_V = 44 \pm 12$ in 
solar units\footnote{Sandage (1986) finds 
$M/L \sim 25$ for $M_{LG} \sim 3 \times 10^{12}$ \msol.  His lower 
$M/L$ estimate is based, in part, on a higher estimate of the total 
luminosity of the Local Group.}.
It is, perhaps, worth noting that M31 and the Galaxy together provide
86\% of the luminosity of the LG.  The uncertainty in $M_V$(Galaxy) 
contributes significantly to the error of the integrated luminosity of 
the Local Group.  

Finally, one can compute the radius of the zero-velocity surface, $r_\circ$,
that separates Hubble expansion from cluster contraction at the 
present epoch (Lynden-Bell 1981, Sandage 1986).  As the universe expands,
the zero-velocity surface moves outward with time.  If the total random
components of the velocity field cancel out, one can write, from Eq. (7)
of Sandage (1986):

\begin{equation}
r_\circ {\rm [Mpc]} = \left( \frac{8GT^2}{\pi^2} M_{\rm LG} \right)^{1/3} 
   = \ 0.154 \ (T[\rm{Gyr}])^{2/3} \ \ (M_{\rm LG}[10^{12} \msol])^{1/3}.
\end{equation}

Assuming that the age of the Local Group is $14\pm2$ Gyr, and using
our estimate of the virial mass of the Local Group, we find
$ r_\circ = 1.18 \pm 0.15 \ \ \rm{ Mpc}. $
The value of $r_\circ$ given above can now be used to assess LG membership.


\section{Local Group membership}

On the basis of the membership criteria listed in \S 1, van den Bergh (1994b) 
concluded that it was safe to exclude the following galaxies from membership 
in the Local Group: (1) the Sculptor irregular (=UKS 2323-326), (2) Maffei 1 
and its companions, (3) UGC-A86 (=A0355+66), (4) NGC 1560, (5) NGC 5237, 
and (6) DDO 187.  A particularly strong concentration of Local Group suspects,
which includes (2),(3),(4) and (5) listed above, occurs in the direction
of the IC 342/Maffei group 
(van den Bergh 1971, Krismer, Tully \& Gioia 1995), which Krismer \etal
place at a distance of $3.6 \pm 0.5$ Mpc.  Cassiopeia 1, regarded as a
Local Group suspect (Tikhonov 1996), also appears to be a member of the 
IC 342/Maffei group.
Van den Bergh \& Racine (1981) failed to resolve Local Group suspect LGS2
on large reflector plates. They conclude that this object is either a
Galactic foreground nebula, or an unresolved stellar system at a much
greater distance than that of M31 and M33.  Another long-time Local Group
suspect is DDO 155=GR 8.  However, observations by Tolstoy \etal (1995)
have resulted in the discovery of a single probable Cepheid, which yields
a distance of 2.2 Mpc, so that this galaxy would lie outside of the
Local Group boundary.  The spiral galaxy NGC 55 has recently been listed as
a possible Local Group member by Mateo (1998).  However, it appears preferable
to follow in the footsteps of de Vaucouleurs (1975), who assigns this galaxy
to the Sculptor (South Polar) group.  C\^ot\'e, Freeman \& Carignan (1994)
show that NGC 55 is located close to the center of the distribution of dwarf
galaxies associated with the South Polar group.  Furthermore photometry in
$J$, $H$,and $K$ by Davidge (1998) shows that NGC 55, NGC 300, and NGC 7793
are located at comparable distances.  Sandage \& Bedke
(1994, panel 318) write ``NGC 55 is very highly resolved into individual
stars, about equally well as other galaxies in the South Polar Group such
as NGC 247 and NGC 300.  Evidently, NGC 55 is just beyond the Local Group.''
Finally, Jergen, Freeman, \& Bingelli (1998) place NGC 55 on the near side 
of the Sculptor group.  We have also excluded the galaxies NGC 3109, Antlia, 
Sextans A, and Sextans B from membership in the Local Group.  These objects, 
which have measured distances of 1.36, 1.70, 1.45 and 1.32 Mpc
respectively (vdB2000), are located relatively close together on the sky.
Their mean distance from the barycenter of the Local Group, which is
situated $\sim 450$ kpc away in the direction towards M~31, is ~1.7 Mpc.
Furthermore these galaxies have a mean redshift of $114 \pm 12$ \kms\ relative
to the $V_r$-$\cos\theta$ relation derived in \S3 (see vdB2000).  These data
suggest that NGC 3109, Antlia, Sex A and Sex B form a physical grouping that
is receding from LG, and that lies just beyond the LG zero-velocity
surface (van den Bergh 1999).  We note that N3109, Sextans A, and 
Sextans B were also excluded
by YTS77 on the basis of their solar motion solutions.
Zilstra \& Minniti (1998) find that the LG candidate IC 5152 is located
at $1.8 \pm 0.2$ Mpc, which places it outside the LG zero-velocity surface.

Group membership can be revised by inspection of the $V_r-\cos\theta$
diagram, which illustrates the motions of individual galaxies with 
respect to the ensemble of the galaxies in the Group.  This is shown 
in Figure 4 for LG galaxies.  The LG radial velocity dispersion, 
$\sigma_r = 61 \pm 8$ \kms, is shown by dotted lines.  Suspected 
outliers lying below the $1\sigma$ regression line are few.  None of 
the systems presented in Fig. 4 can be excluded from membership on the 
basis of this test.  The two blue-shifted systems (IC 1613 and Pisces), and 
handful of redshifted LG objects, fall within 2$\sigma$ of the regression
line.  Membership for many recently discovered dwarf Spheroidals cannot 
be examined with this test, because their radial velocities are not yet 
available.  

Figure 1 shows that most of the LG members are concentrated in subgroups
that are centered on the Andromeda galaxy and on the Milky Way system.
However, a few objects, such as NGC 6822, IC 1613, Leo A, and the WLM 
system, appear to be free-floating Group members.  Aquarius (=DDO 210),
Tucana, and SagDIG are so far from the barycenter of the LG that their 
membership in the Local Group cannot yet be regarded as firmly
established, even though they lie close to the solar ridge-line 
in the $V_r-\cos\theta$ diagram. 

It might be argued that our value of $\sigma_r$ is biased low because 
the data base may lack (unknown) nearby fast-moving galaxies.  However, 
this effect is probably not important because no galaxy are found 
with large blue-shifts relative to the mean relationship 
between $\cos\theta$ and apex distance.

\section{Discussion and Summary}

We have measured a new solution for the solar motion relative to LG
galaxies which agrees very well with previous derivation by, \eg
YTS77, Sandage (1986), and Karachentsev \& Makarov (1996).  Following
RG98 it is worth pointing out that these solutions are only physically
meaningful under the assumption that the 3-dimensional spatial and 
velocity distributions are independent.  This would not be true if 
the LG potential is bimodal.  This is verified by computing the motion
of the Sun relative to the two main LG substructures, the Andromeda 
and Galaxy subgroups, and testing if their combined motion matches 
that inferred relative to the entire Local Group.  Preliminary 
indications suggest that the Andromeda subgroup is moving faster 
with respect to the Sun, and in a different direction from the 
Galaxy subgroup. However, the error bars (mostly for 
$V_{Sun \rightarrow Andsub}$) are far too large to rule out 
the ``standard'' solution.  Indeed, the combined subgroup 
solutions are perfectly consistent with our final derivation for 
the solar motion relative to all LG members with 
$V_{{\rm Sun} \rightarrow {\rm LG}} = 306\pm 18$ \kms\ towards an 
apex at $\ell= 99^\circ \pm 5^\circ$ and $b = -4^\circ \pm 4^\circ$.
It is worth pointing out that interpretation errors for the solar motion 
may linger until we get a better understanding of the true orbital 
motions of LG members, and a better knowledge of the overall 
distribution of mass in the Local Group.

The observed radial velocity dispersion of the LG is 
$\sigma_r = 61 \pm 8$ \kms.
Braun \& Burton (1998) measured a radial velocity dispersion 
$\sigma_r= 69$ \kms\ for the motion of intra-cluster compact \hi 
high-velocity clouds (HVCs).  Although close in dispersion to the 
LG value, HVCs exhibit an excess of infall velocities (blue-shifts) 
suggesting that many of them may presently still be falling into the 
Local Group, contrary to bona fide LG galaxies. 

Table 1 presents an updated listing of 35 probable members of the 
Local Group.  Half of all the members are located within 450 kpc 
of the barycenter of the LG, with only 3 objects,
SagDIG, Aquarius, and Tucana, being more than 1 Mpc away. 
These results show that the (binary) core of the LG is relatively
compact, and well-isolated from other nearby clusters.  
This conclusion was already anticipated by Hubble in his 
``Realm of the Nebulae'' (1936): ``the Local Group is [a] typical, 
small group of nebulae which is isolated in the general field.''
One must, however, remain cautious about these statements in light 
of the recent, surprisingly high, rate of discoveries of new
members that are low surface brightness dwarfs.  These have all
been discovered at distances $\ll r_\circ$, for obvious 
observational reasons, but it would be premature to exclude a 
significant population of low surface brightness galaxies at 
greater distances as well.

Adopting a half-mass radius $r_h = 0.35$ Mpc and a LG age of
$14\pm2$ Gyr yields a radius $r_\circ= 1.18 \pm 0.15$ Mpc for the 
zero-velocity surface of the Local Group, and a total LG mass 
$M_{LG}=(2.3\pm0.6) \times 10^{12}$ \msol.  This mass determination
is, of course, only valid {\it if} the LG is in virial equilibrium.  
The fact that an equal number of LG members are blue-shifted and 
redshifted relative to the adopted solar motion suggests that the 
LG may be {\it at least} approaching virial equilibrium.  An 
independent ``projected'' mass estimate for the Andromeda subgroup,
combined with published mass information for the Galaxy subgroup by 
Zaritsky (1999), yield nearly the same total mass for the LG,
independent of any assumption about the virial nature of the LG.
With this mass, the visual mass-to-light ratio (in solar units) 
for the LG is $44 \pm 12.$

Mulchaey \& Zabludoff (1998) find that the velocity dispersion of 
clusters of galaxies, and their X-ray luminosities and temperatures,
are related by the relations; 

\begin{equation}
\log L_X = 31.6 \pm 1.1 + \log h^{-2} + (4.3 \pm 0.4) \log \sigma_r, 
\end{equation}

and

\begin{equation}
\log L_X = 42.44 \pm 0.11 + \log h^{-2} + (2.79 \pm 0.14) \log T , 
\end{equation}

\noindent 
where the dimensionless Hubble ratio $h$, is given by \hub = 100 h\hunit.
Extrapolating these relations to small values of $\sigma_r,$ and adopting
$h=0.65$ and $\sigma_r = 61 \pm 8$ \kms, one obtains $T \approx 74$ eV and 
$L_X \approx 4.5 \times 10^{39}$ erg s$^{-1}$ for the intra-cluster gas in 
the Local Group.  These numbers suggest that it would be difficult, with 
current X-ray instrumentation and due to the strong absorption by our galaxy
below 0.5 keV, to detect X-ray emission from any small group like the LG.  

{\noindent {\bf Acknowledgements}}

We would like to thank St\'ephane Rauzy for sharing his likelihood
estimator and useful comments on the paper.

\clearpage


\begin{figure}
\epsscale{1.00}
\caption{Positions of Local Group members in Galactic cartesian 
 coordinates, as viewed from two orthogonal directions.  The left side
 shows a sphere of radius $r_\circ = 1.18$ Mpc (corresponding to the
 zero-velocity surface of the Local Group) as a solid line.  
 The dotted-line shows a sphere of radius $r_h = 450$ kpc which 
 encompasses both the Andromeda and Galaxy subgroups.  Both spheres 
 are centered on the LG barycenter at X=-220, Y=+361, and Z=-166 kpc. 
 For clarity, not all the LG members have been labeled.
 Filled dots represent galaxies within the Andromeda/Galaxy volume; 
 LG members outside of this sphere are plotted as crosses.  The Galaxy, 
 M31, and M33 are shown with a spiral galaxy symbol.
}
\end{figure}

\begin{figure}
\epsscale{1.00}
\caption{Histogram showing the distribution of measured distances of
 all known LG members from the LG dynamical center.  It is 
 seen that the membership drops steeply beyond $D_{LG} \simeq 0.85$ Mpc.  
 The density of galaxies near D$_{\rm LG} =0$ is small because the center 
 of the LG is situated between the Andromeda and Galactic subgroups, where
 few galaxies are found.
}
\end{figure}

\begin{figure}
\epsscale{1.00}
\caption{Cumulative distance distribution of all known LG members.  
This histogram shows that the core of the Local Group has $\ltorder 0.85$ Mpc.
Half of the known members of the Local Group are seen to located within 450 kpc 
of the adopted barycenter. 
}
\end{figure}

\begin{figure}
\epsscale{1.00}
\caption{Aitoff projection in Galactic coordinates showing the residuals
 between our solution for the solar motion relative to the Local Group 
 (Eq. 6) and that of Rauzy \& Gurzadyan (1998; Eq. 7). 
 The size of the symbols is linearly proportional to the magnitude 
 of the residual, the largest ones being +154 \kms\ in the direction 
 $\ell = 145^\circ, b=60^\circ$, and -154 \kms\ toward $\ell =325^\circ, 
 b=-60^\circ$.  Positive and negative residuals (This paper - RG98) are 
 shown with circles and stars, respectively. 
}
\end{figure}

\begin{figure}
\epsscale{1.00}
\caption{Observed heliocentric velocities $V_r$ of Local Group
members versus $\cos\theta$, where $\theta$ is the angular distance 
from the solar apex.  Our solar motion solution of  $306 \pm 18$ \kms\
towards $\ell =99^\circ \pm 5^\circ$ and $-3^\circ \pm 4^\circ$ is shown
as the solid ridge line.  Dotted lines correspond to the residual
radial velocity dispersion of $\pm61$ \kms\ from the ridge solution. 
Note the large deviations of Leo I and of the Sagittarius dwarf
(which is strongly interacting with the Galaxy) from the mean motion
of Local Group members.
}
\end{figure}

\begin{figure}
\epsscale{1.00}
\caption{Same as Figure 5 but using the solar motion solution of 
Rauzy \& Gurzadyan (1998) with V\sol = $306 \pm 18$ \kms\
towards $\ell =94^\circ \pm 48^\circ$ and $-34^\circ \pm 29^\circ$.
For comparison, the solid ridge line and dotted dispersion lines 
are those computed for Figure 5. 
Note that the dispersion around the regression relation of Rauzy
\& Gurzadyan (1998; Eq. 7) is twice larger ($\sigma_r = 110.3$ \kms)
than that for the Standard solution shown in Figure 5.
}
\end{figure}

\begin{figure}
\epsscale{1.00}
\plotone{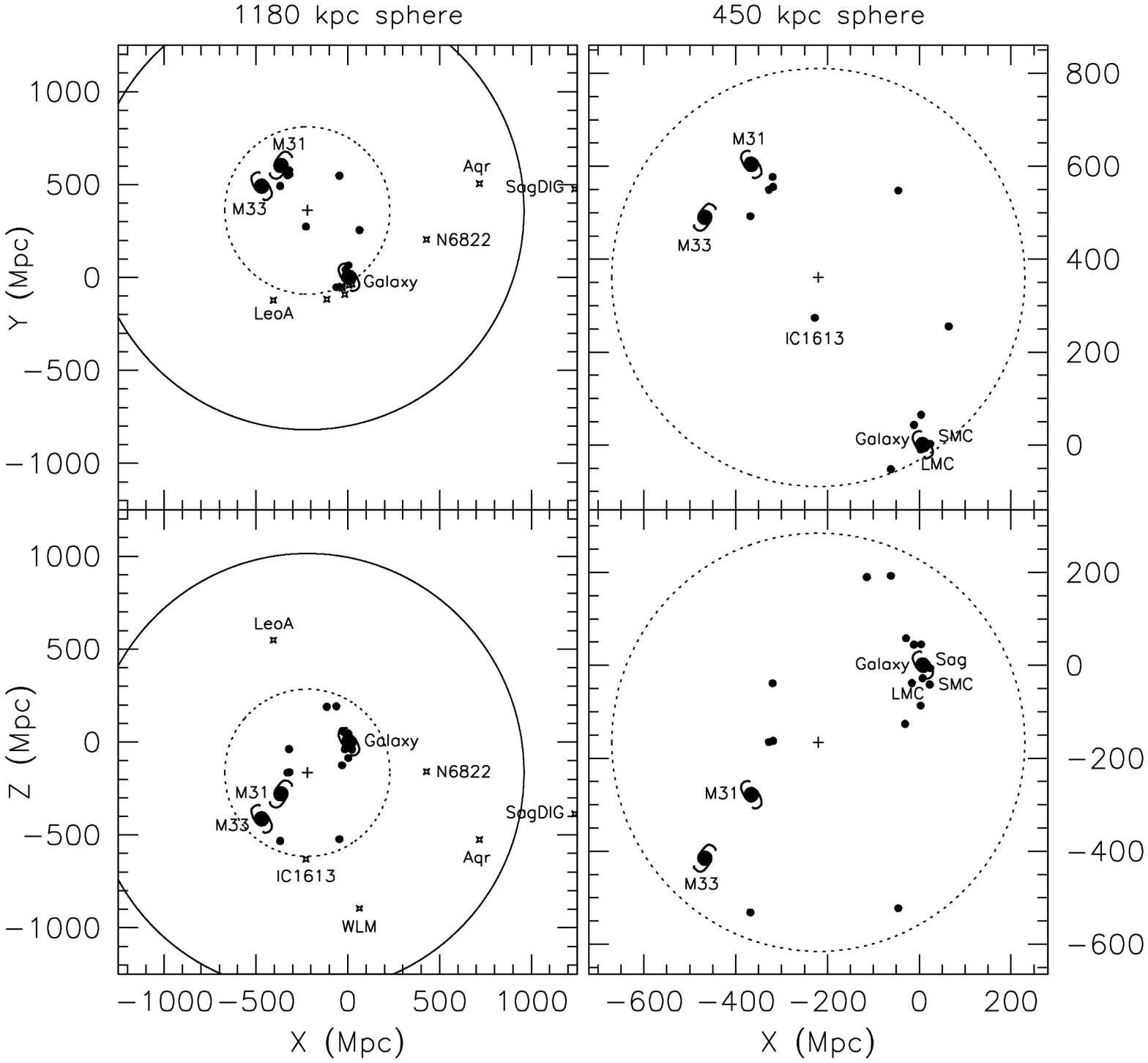}
\end{figure}
\clearpage

\begin{figure}
\epsscale{1.00}
\plotone{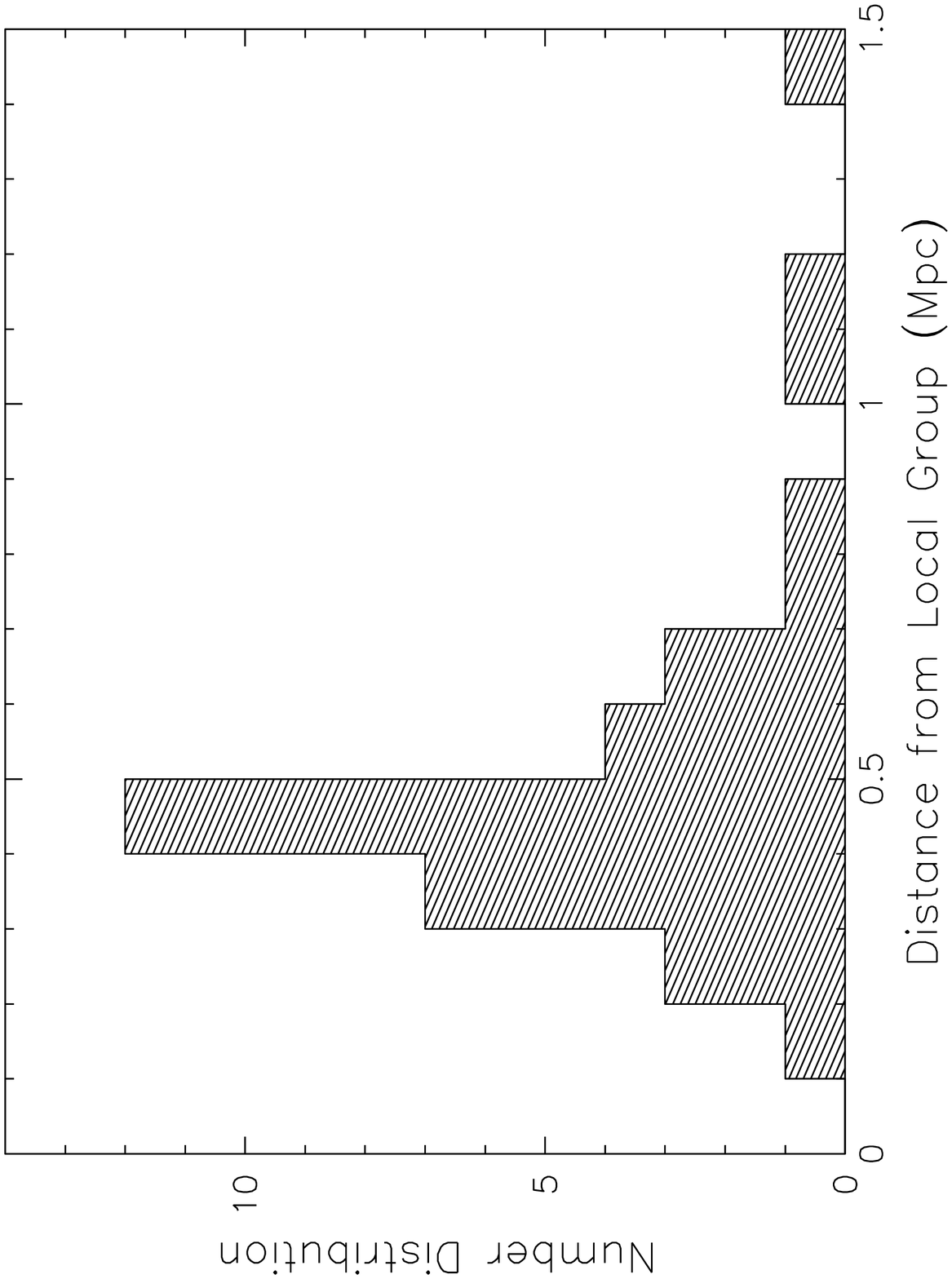}
\end{figure}

\begin{figure}
\epsscale{1.00}
\plotone{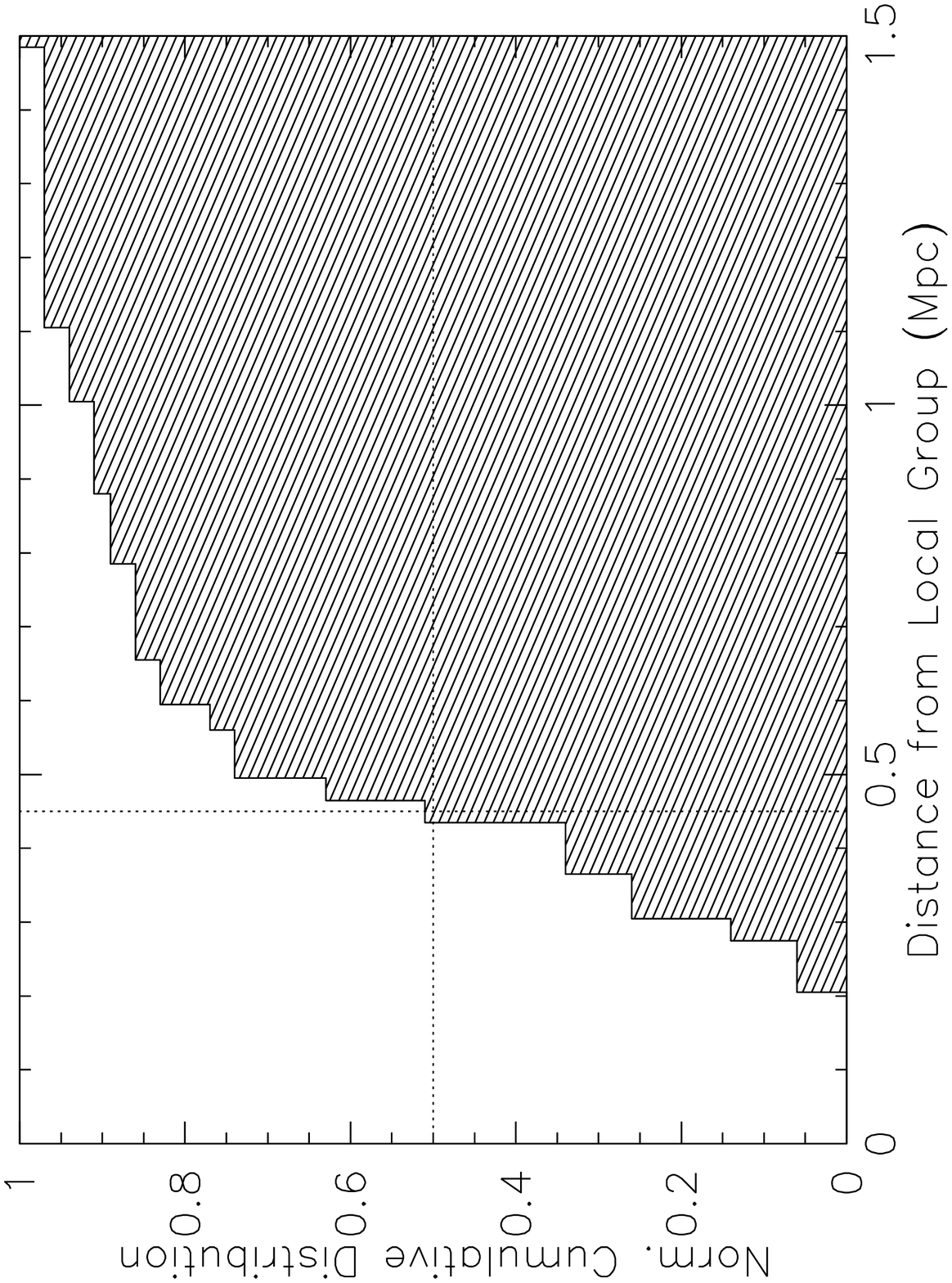}
\end{figure}

\begin{figure}
\epsscale{1.00}
\plotone{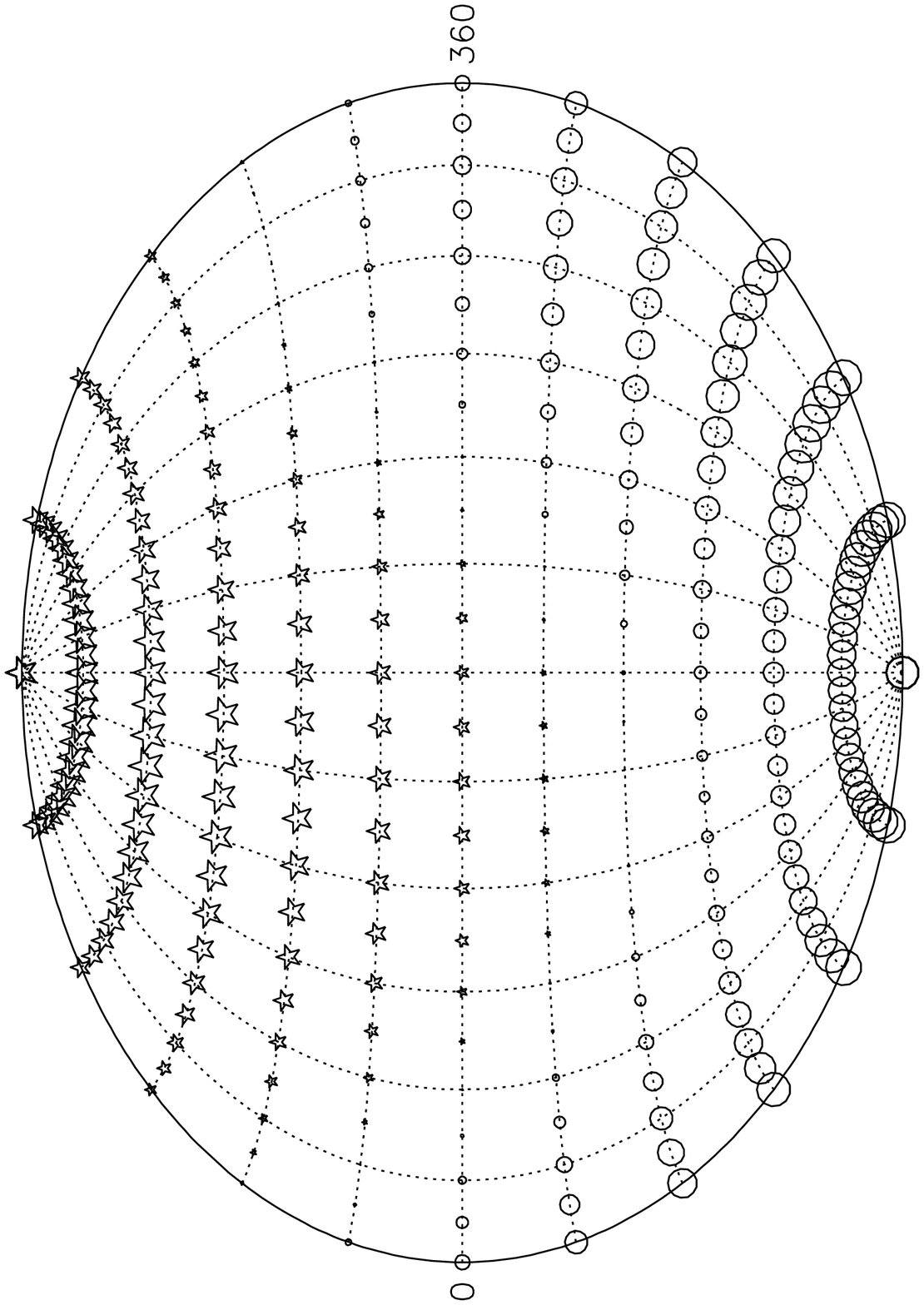}
\end{figure}
\clearpage

\begin{figure}
\epsscale{1.00}
\plotone{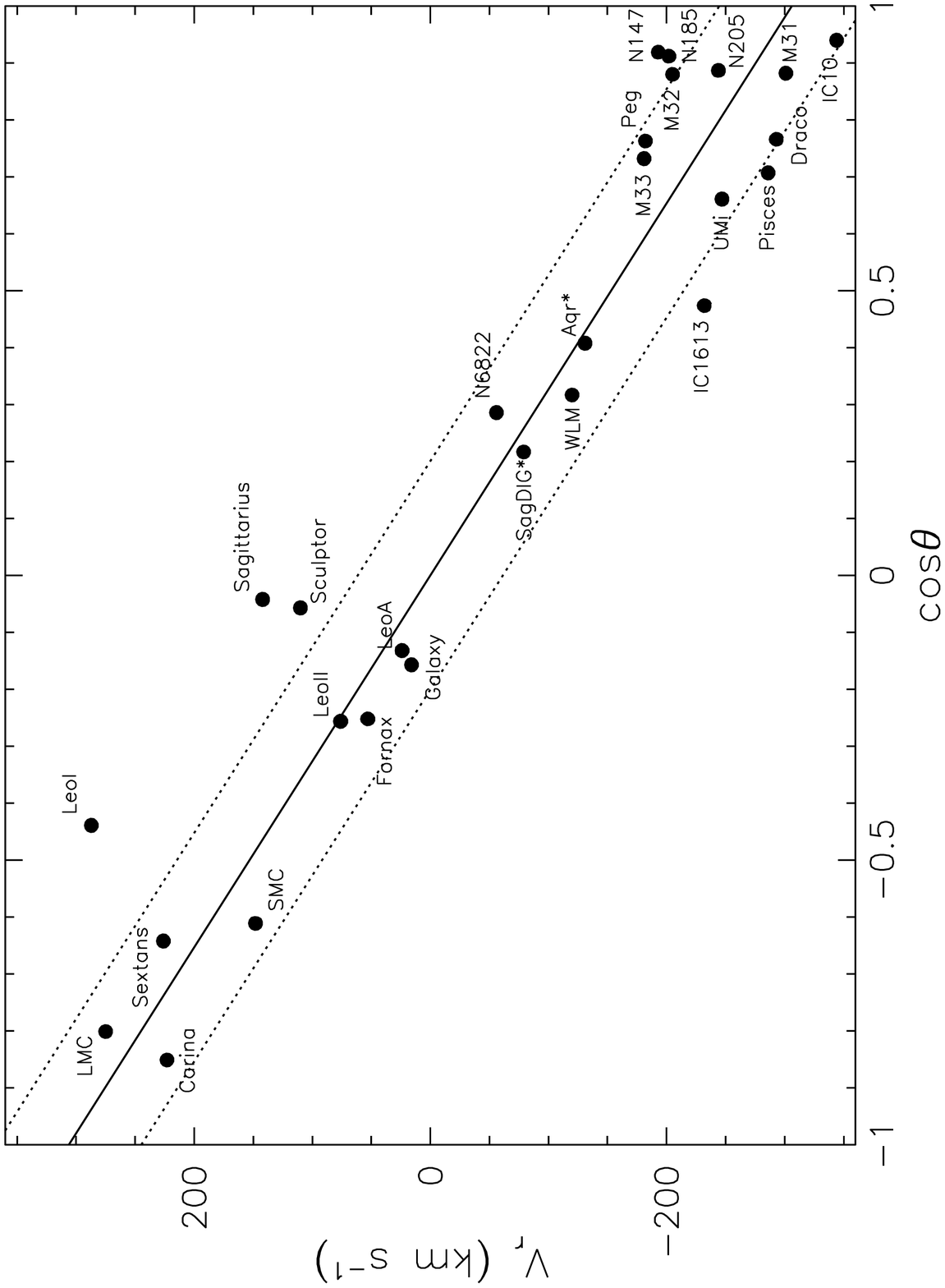}
\end{figure}

\begin{figure}
\epsscale{1.00}
\plotone{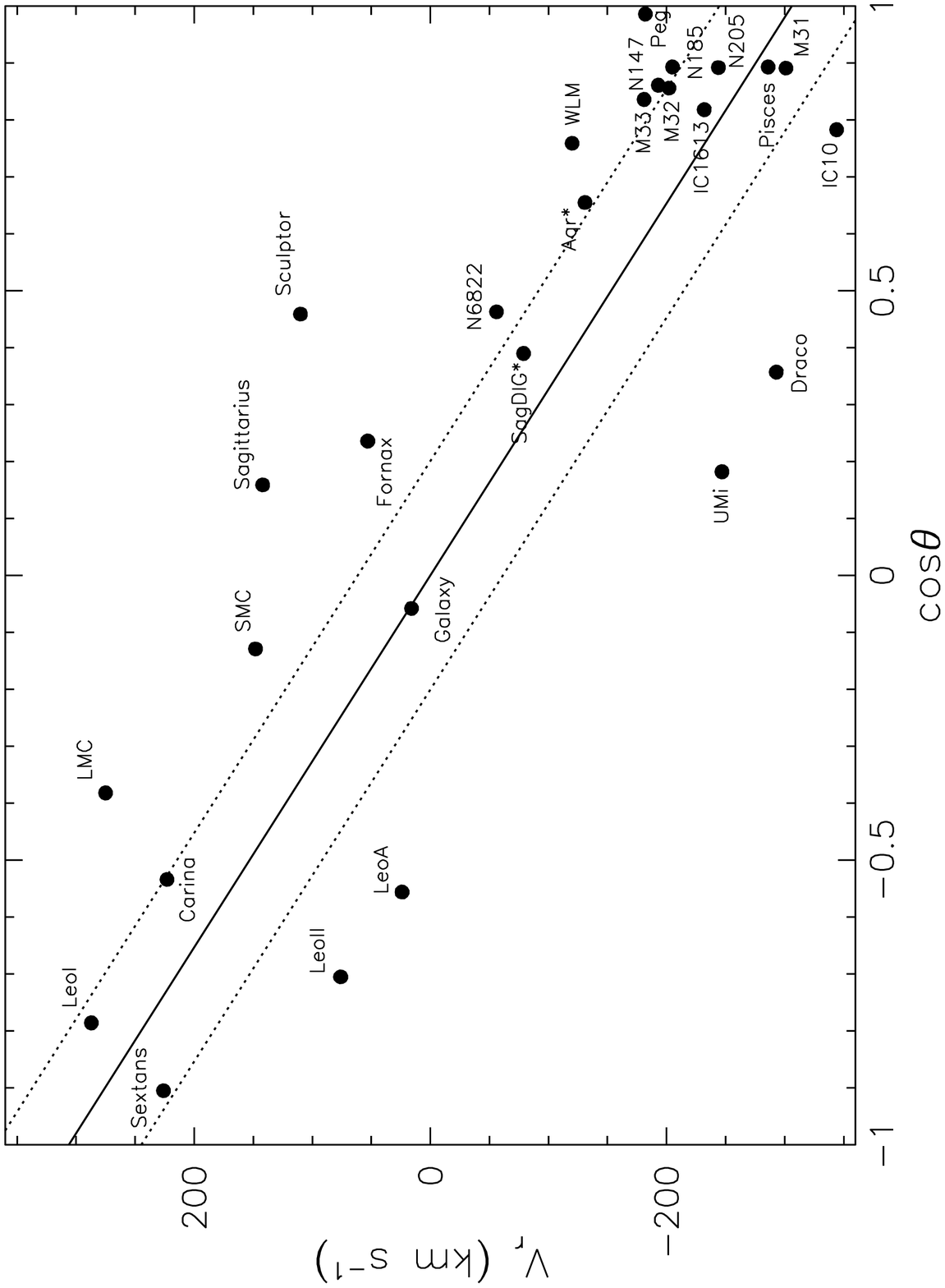}
\end{figure}

\end{document}